\documentclass[floatfix,showpacs,showkeys,twocolumn,preprintnumbers,amsmath,amssymb]{revtex4}
\usepackage{graphicx}
\usepackage{dcolumn}
\usepackage{amsmath}
\usepackage{subfigure,hyperref}
\usepackage[normalem,normalbf]{ulem}

\newcommand{\be}{\begin{equation}}
\newcommand{\ee}{\end{equation}} 
\newenvironment{Eqnarray}%
                   {\arraycolsep 0.14em\begin{eqnarray}}{\end{eqnarray}}

\newcommand{\half}{\text{$\frac{1}{2}$}} 
\newcommand{\Tr}{\text{Tr}}
\newcommand{\p}{\partial} 
\newcommand{\bpsi}{\bar{\psi}} 
\newcommand{\bphi}{\bar{\phi}}

\def\nbC{{\mathchoice {\setbox0=\hbox{$\displaystyle\rm C$}%
\hbox{\hbox to0pt{\kern0.4\wd0\vrule height0.9\ht0\hss}\box0}} 
{\setbox0=\hbox{$\textstyle\rm
C$}\hbox{\hbox to0pt{\kern0.4\wd0\vrule height0.9\ht0\hss}\box0}} 
{\setbox0=\hbox{$\scriptstyle\rm
C$}\hbox{\hbox to0pt{\kern0.4\wd0\vrule height0.9\ht0\hss}\box0}}
{\setbox0=\hbox{$\scriptscriptstyle\rm C$}\hbox{\hbox to0pt{\kern0.4\wd0\vrule
height0.9\ht0\hss}\box0}}}}
\def\nbQ{{\mathchoice {\setbox0=\hbox{$\displaystyle\rm 
Q$}\hbox{\raise 0.15\ht0\hbox
to0pt{\kern0.4\wd0\vrule height0.8\ht0\hss}\box0}} 
{\setbox0=\hbox{$\textstyle\rm Q$}\hbox{\raise
0.15\ht0\hbox to0pt{\kern0.4\wd0\vrule height0.8\ht0\hss}\box0}} 
{\setbox0=\hbox{$\scriptstyle\rm
Q$}\hbox{\raise 0.15\ht0\hbox to0pt{\kern0.4\wd0\vrule 
height0.7\ht0\hss}\box0}}
{\setbox0=\hbox{$\scriptscriptstyle\rm Q$}\hbox{\raise 0.15\ht0\hbox 
to0pt{\kern0.4\wd0\vrule
height0.7\ht0\hss}\box0}}}}
\def\nbT{{\mathchoice {\setbox0=\hbox{$\displaystyle\rm 
T$}\hbox{\hbox to0pt{\kern0.3\wd0\vrule
height0.9\ht0\hss}\box0}} {\setbox0=\hbox{$\textstyle\rm 
T$}\hbox{\hbox to0pt{\kern0.3\wd0\vrule
height0.9\ht0\hss}\box0}} {\setbox0=\hbox{$\scriptstyle\rm 
T$}\hbox{\hbox to0pt{\kern0.3\wd0\vrule
height0.9\ht0\hss}\box0}} {\setbox0=\hbox{$\scriptscriptstyle\rm T$}\hbox{\hbox
to0pt{\kern0.3\wd0\vrule height0.9\ht0\hss}\box0}}}}
\def\nbS{{\mathchoice {\setbox0=\hbox{$\displaystyle     \rm 
S$}\hbox{\raise0.5\ht0%
\hbox to0pt{\kern0.35\wd0\vrule height0.45\ht0\hss}\hbox 
to0pt{\kern0.55\wd0\vrule
height0.5\ht0\hss}\box0}} {\setbox0=\hbox{$\textstyle        \rm 
S$}\hbox{\raise0.5\ht0%
\hbox to0pt{\kern0.35\wd0\vrule height0.45\ht0\hss}\hbox 
to0pt{\kern0.55\wd0\vrule
height0.5\ht0\hss}\box0}} {\setbox0=\hbox{$\scriptstyle      \rm 
S$}\hbox{\raise0.5\ht0%
\hboxto0pt{\kern0.35\wd0\vrule height0.45\ht0\hss}\raise0.05\ht0%
\hbox to0pt{\kern0.5\wd0\vrule height0.45\ht0\hss}\box0}} 
{\setbox0=\hbox{$\scriptscriptstyle\rm
S$}\hbox{\raise0.5\ht0%
\hboxto0pt{\kern0.4\wd0\vrule height0.45\ht0\hss}\raise0.05\ht0%
\hbox to0pt{\kern0.55\wd0\vrule height0.45\ht0\hss}\box0}}}}
\def\nbZ{{\mathchoice {\hbox{$\sf\textstyle Z\kern-0.4em Z$}} 
{\hbox{$\sf\textstyle Z\kern-0.4em Z$}}
{\hbox{$\sf\scriptstyle Z\kern-0.3em Z$}} 
{\hbox{$\sf\scriptscriptstyle Z\kern-0.2em Z$}}}}

\makeatletter
\renewcommand{\@thesubfigure}{\alph{subfigure}) \space}
\renewcommand{\p@subfigure}{}
\makeatother
\begin{document}
\title{What can be learnt from the nonperturbative renormalization group?}

\author{B. Delamotte} \email{delamotte@lptl.jussieu.fr}
\affiliation{Laboratoire de Physique Th\'eorique des Liquides, Universit\'e Paris VI-Pierre et Marie Curie, 75252 Paris Cedex 05, France.  }%
\author{L. Canet} \email{leonie@theory.ph.man.ac.uk}
\affiliation{School of Physics and Astronomy, University of Manchester, Manchester M13 9PL, United Kingdom.}%

\date{\today}

\begin{abstract}
We point out some limits of the perturbative renormalization group  used in statistical mechanics both
at and out of equilibrium. We argue that the non perturbative renormalization group formalism is a promising candidate to overcome
some of them. We present some results recently obtained
 in the literature that substantiate our claims. We finally list some open issues for which this formalism could be useful 
and also review some of its  drawbacks.
\end{abstract}

\maketitle

\section{Introduction}
Field theory together with renormalization group (RG)  have provided  very powerful
means to investigate continuous phase transitions in equilibrium and non equilibrium statistical physics.
The need for field theoretical techniques is now well understood: when spatial (and/or time) fluctuations
are large  in a $N$-body system, the mean field approach is inadequate and it becomes necessary to keep track
of the space (and/or time) dependence of the order parameter. In the continuum, statistical fluctuations are summed
over through a functional integral so that  theories  to be dealt with are (euclidean) field theories.
The difficulty with these theories is that close to a continuous phase transition, fluctuations on all length
 (and/or time) scales contribute and render the perturbative approach problematic. The origin of the problem
is  well known and is twofold. First, at each order of the perturbation expansion (of, say, 
a correlation function) all fluctuations of all wave lengths (and/or frequencies) contributing to this order
are summed over (Feynman graphs). Since they all contribute algebraically, the resulting sums (integrals) diverge at very short 
(ultraviolet) or very long (infrared) wavelengths. Getting rid of these ultraviolet divergences is the subject of perturbative
renormalization and is now well under control\cite{lebellac92,goldenfeld92,itzykson89}. Second, even after renormalization, the 
perturbation series are in general non convergent and, as such, difficult to use to obtain reliable quantitative
results. This is, for instance, the case for the perturbation expansion of the $\beta$-functions of the RG flow
and of the critical exponents. This is all the more severe  that the  dimensions of interest are far below  the 
critical dimension since, then, the field theories are in their strong coupling regime. Fortunately,
in some cases, the perturbation series turn out to be Borel summable so that efficient resummation techniques
(Pad\'e-Borel, conformal mappings, etc) allow to compute physical quantities reliably and with high accuracy.
This is the case for the $O(N)$ models in three dimensions for which five ($\epsilon$-expansion)  and
six (fixed dimension expansion) loops have been computed and where all available methods lead to consistent and
accurate results for critical exponents\cite{zinnjustin89}. Some of the most celebrated equilibrium systems for which resummation
techniques also work well are ferromagnetic systems with cubic anisotropy or with quenched disorder\cite{pelissetto00a,holovatch02,folk03}. However, and contrary to the common belief, this ideal picture turns out
to be the exception rather than the rule.
In most cases, the situation is more complicated for at least two reasons. First, in many cases, only the two first
orders of perturbation series are known and are insufficient to perform any resummation. This is the generic 
case in out of equilibrium statistical systems and for equilibrium ones involving fermions or gauge fields (as, for instance,
the electromagnetic field). Second, even in cases where five or six loop series are known, resummations do not always lead to converged
results. This is the case, for instance, for frustrated spin systems in three dimensions\cite{tissier00,pelissetto01a,pelissetto02,tissier03,delamotte04,calabrese04}.
 Again, contrary to the common belief, the problem is not just to compute accurately
critical exponents but to determine the {\it qualitative} behavior of the system at the transition: either first or second
order phase transition, belonging or not to a given universality class, etc. 

Of course, apart from these difficulties,
more serious ones --- not only related to  strong coupling behaviors --- can exist and  invalidate perturbation expansions. 
Many examples exist in different physical
contexts. The low temperature expansion of the ferromagnetic XY ($O(2)$) model in two dimensions is identical at all orders
to that of a free theory. This expansion is invalidated at finite temperature by the existence and liberation of vortices
(Kosterlitz-Thouless transition)\cite{itzykson89}. Quantum Chromodynamics (QCD) is unable at any finite order of perturbation to describe
 quark confinement. The perturbation expansion of the $\beta$-function of the Kardar-Parisi-Zhang (KPZ) equation describing 
the growth of interfaces is known at all orders (it is given by the one-loop result) but is unable to describe the 
scaling behaviors in the rough phase above two dimensions\cite{frey94,wiese98}. The $\epsilon=6-d$ expansion of the random field $O(N)$ model
(RFO(N)M) is identical at all orders to the $\epsilon=4-d$ expansion of the pure model (dimensional reduction)\cite{parisi79}. However, it has been
 rigorously proven that the RFIM in $d=3$ and the pure Ising model in $d=1$ display different critical behaviors\cite{imbrie84,bricmont87}. 
This implies that dimensional reduction must break down, at least below a dimension between six and three.

In the two first examples, the problem is that the low energy excitations relevant for the description of the physics at
large scale are qualitatively different from those at small scale: they are vortices for the XY model and bound 
states (actually confined states) like hadrons for QCD. Since, on the one hand, these excitations dominate
the low energy physics and, on the other hand, they  lead to non analytic contributions that
 are missed by  perturbation expansions, the latters are unable to describe the large scale physics of these
systems. The presence of non analyticities is also the reason for the breakdown of dimensional reduction in the
RFO(N)M\cite{tarjus05}. It is not known, although suggested in the literature, whether these non analyticities are also related to 
the presence of a bound state\cite{brezin01,parisi02}, and neither it is for the KPZ equation.

All these difficulties suggest that it is necessary to go beyond perturbation theory, at least to be able to deal with strongly 
coupled systems without having to resort to perturbation expansions at high
orders and to resummations. Unfortunately, apart from two dimensions or at large $N$, no systematic method for solving exactly a field
theory is known. One has to go back to the summation over the fluctuations in the partition function and try to organize it 
differently. This is what Wilson did with the momentum shell integration of rapid modes and the construction of effective theories
for slow modes\cite{wilson74}. His idea was to integrate out fluctuations scale by scale and not order by order in a series expansion. 
Here, an important remark is of order. If it were possible to perform exactly
this integration between two different scales, it would be possible to iterate this procedure until all fluctuations are integrated
out. This would amount to solving exactly the problem and, of course, this is impossible in general. Actually, there are only very 
few examples where RG techniques have enabled to solve exactly a model that was not solved (in a simpler way) by another method.
In a sense, Wilson's procedure is more subtle. Of course, the momentum shell integration can be performed exactly in a formal 
way, that is by expressing the RG flow of a quantity --- e.g. an effective hamiltonian --- in terms of itself. This leads
to the formulation of an exact RG equation (that we derive in the following and which is a functional differential 
equation)\cite{wilson74,polchinski84}. But this equation cannot be solved in general.
The important point in Wilson's method is that the very idea behind it suggests a natural way to approximate the integration
over the rapid modes and thus a natural truncation of the exact RG equation. Thus, RG techniques are almost intrinsically
linked with approximation methods: they are not very interesting when exact solutions are known.

The aim of this article is to show how Wilson's RG ideas can be concretely implemented both at and out of equilibrium.
In particular, we shall show how  Wilson's idea of a flow of effective hamiltonians is conveniently replaced by that of a flow 
of Gibbs free energies and we shall generalize it to non equilibrium statistical mechanics where the free energy no longer exists.
One of the crucial part of this article will, of course, be devoted to the most developed truncations of this flow. In particular, we 
shall show that they  allow both to  accurately determine universal quantities, such as critical exponents even when the theories are strongly
coupled, and non universal quantities, such as phase diagrams. This last point is a crucial advantage of this method as compared with 
perturbation theory.

\section{The non perturbative renormalization group (NPRG) formalism}

\subsection{The equilibrium case}
Any implementation of Wilson's momentum shell integration relies on a separation between rapid and slow modes. This separation can be 
 achieved through a sharp cut-off in momentum space but this procedure leads to singularities  when 
 approximations are performed. It is preferable to perform a smooth separation by modifying the Boltzmann weights in such a way
that the slow modes effectively decouple from the model while the rapid ones are unaffected. This is conveniently achieved by adding 
to the original partition function (for equilibrium systems) a ``momentum-dependent mass term'' giving a large ``mass'' to slow modes
and leaving unchanged the rapid ones. The slow modes are thus weakly correlated, they do no longer propagate and do not contribute
to the long distance physics of the (modified) model. For a theory with one scalar field $\phi$ and hamiltonian $H[\phi]$, we hence define 
a scale dependent family of partition functions\cite{wetterich93c,tetradis94,ellwanger94c,morris94a,morris94b,berges02}:
\be
{\cal Z}_k[B] = \int \: {\cal D}\phi\;  e^{
-H[\phi]- \Delta H_k[\phi] +\int B.\phi}, 
\label{zk}
\ee
where $B$ is an external source, e.g. a magnetic field, $k$ a momentum scale  and $\Delta H_k[\phi]$ is the scale 
dependent mass term responsible for the decoupling of slow modes  in ${\cal Z}_k$:
\begin{Eqnarray}
\Delta H_k[\phi] &=& \half \int d^dx\: R_k(x-y)\phi(x)\phi(y)\\
        &=&  \half \int \frac{d^dq}{(2\pi)^d}\: R_k(q)\phi(q)\phi(-q).
\end{Eqnarray}
  $R_k$ is the cut-off function that must verify:
\begin{Eqnarray}
q^2\gg k^2&:& \ R_k(q^2)\to0\ \ \ \hbox{no mass for rapid modes}\\
q^2\ll k^2&:&\ R_k(q^2)\sim k^2\ \ \hbox{large mass for slow modes,}\ \ 
\end{Eqnarray}
in order to achieve the decoupling. A typical and very simple function satisfying these constraints is\cite{litim01}
\be
R_k(q)= (k^2-q^2)\theta(k^2-q^2)
\label{regulateur}
\ee
which is a parabola with negative concavity on $[0,k]$ and vanishes elsewhere. These definitions yield that: 

1) when $k=0$, $R_{k=0}(q)=0$ identically and thus ${\cal Z}_{k=0}={\cal Z}$, the original partition function  of the model
under study. The latter is therefore recovered in the limit $k\to 0$, that is when all fluctuations have been integrated out.

2) when $k=\Lambda\sim$ inverse lattice spacing (or, more generally, the microscopic scale), $R_k(q)$ is very large for all modes
so that all fluctuations are frozen. Thus
\be
\log {\cal Z}_{k=\Lambda} \sim -H[M^{ MF}] -\Delta H_{k=\Lambda}[M^{MF}] +\int B M^{MF}
\label{champmoyen}
\ee
where $M^{MF}$ is the configuration with highest Boltzmann weight, that is $M^{MF}$ is the mean field magnetization of the
model with hamiltonian $H +\Delta H_{\Lambda}$. 

Then, if one defines the family of functionals $\Gamma_k[M]$  given by the Legendre transform 
of the free energy ${\cal W}_k[B]=\log {\cal Z}_k[B]$ (up to the last term proportional to $R_k$)\cite{berges02}:
\be
\Gamma_k[M] + {\cal W}_k[B]= \int B M -\half \int_q R_k(q^2) M_q M_{-q}
\label{legendre}
\ee
with
\be
M(x)= \frac{\delta {\cal W}_k}{\delta B(x)},
\label{aimantation}
\ee
it follows that:

1) when $k=0$, $\Gamma_{k=0}= \Gamma$ is the Gibbs free energy of the system since it becomes in this limit the standard Legendre
transform of $W$ ($R_{k=0}$ is identically vanishing);

2) when  $k=\Lambda$, $\Gamma_{k=\Lambda}[M]\sim H[M]$ from Eqs.(\ref{champmoyen}) and (\ref{legendre}).

\noindent{Thus}, $\Gamma_k[M]$ smoothly interpolates  between the microscopic hamiltonian and the Gibbs free energy when $k$ is decreased from
$\Lambda$ to 0, that is, when more and more flucuations are integrated out\cite{berges02}. The interpretation of $\Gamma_k$ and $M$ is simple.
When $B(x)$ is kept fixed, $M$ as defined by Eq.(\ref{aimantation}) is $k$-dependent. It is the precursor at scale $k$ of the
physical magnetization  and corresponds to the magnetization of a system composed of blocks of spins of size $k^{-1}$, that is
 of a system where only fluctuations inside blocks of size $k^{-1}$ have been summed over.  $\Gamma_k[M]$ is the Gibbs free energy for 
the configuration with given magnetization $M$ of this system. Thus, contrary to Wilson's effective hamiltonian at scale $k$,
which is the hamiltonian for the slow modes that have not yet been integrated out in ${\cal Z}$,  $\Gamma_k[M]$ is the free energy for the rapid modes that have already been integrated out. It enables, in principle, to compute correlation functions of the rapid modes, contrary
to Wilson's effective hamiltonians. $\Gamma_k[M]$ is called the effective average action ($\Gamma_{k=0}[M]$ is called the effective action 
in the context of quantum field theory; it is the generating functional of the 1-Particle-Irreducible (1PI) correlation functions). 

The exact evolution equation of $\Gamma_k[M]$ with $k$ is obtained by differentiating Eq.(\ref{zk}) with respect to $k$ and by eliminating 
${\cal W}_k$ for  $\Gamma_k[M]$ thanks to Eq.(\ref{legendre}). It writes\cite{berges02}:
\be
\p_s\Gamma_k = \half \int_q \dot{R}_k(q^2) (\Gamma^{(2)}_k(q,-q)+R_k(q^2))^{-1}
\label{rgexact}
\ee
where $s=\log(k/\Lambda)$, $\dot{R}_k=\p_sR_k$,
\be
\Gamma^{(n)}_k(q_1,\dots,q_n,M)= \frac{\delta^n\Gamma_k[M]}{\delta M_{q_1}\dots\delta M_{q_n}}
\ee
and $(\Gamma^{(2)}_k(q,-q)+R_k(q^2))^{-1}$ is the inverse, in the operator sense, of $\Gamma^{(2)}_k(q,-q,M)+R_k(q^2)$. Eq.(\ref{rgexact}) is exact and, as such,
contains all perturbative and non perturbative physics. It is a functional (since $\Gamma^{(2)}_k$ depends on $M$), 
partial differential equation that requires truncations
to be solved. The very idea of incomplete integration and effective theories is to preserve as much as possible the long distance
physics, approximating the short distance one. Thus, it is natural to expand $\Gamma_k[M]$ in powers of  derivatives of $M$. This is
the so-called derivative expansion. In practice, this amounts to proposing an ansatz for $\Gamma_k[M]$ which is polynomial in the derivatives
of the order parameter. In the case of a one component scalar field theory with $\nbZ_2$ symmetry, this consists in considering for 
instance\cite{tetradis94,berges02}:
\be
\Gamma_k[M]= \int d^dx\; \left\{U_k(\rho)+ \half Z_k(\rho)(\nabla M)^2 + O(\nabla^4) \right\}
\label{ansatz}
\ee
where $\rho= M^2/2$, $U_k$ and $Z_k$ (that should not be confused with the partition function ${\cal Z}_k$) are functions of $M^2$ and $k$. 
For uniform magnetization, $\Gamma_k[M]$ reduces to $U_k$ (up to a volume factor) which is called the effective potential. The term 
proportional to $Z_k(\rho)$ is
the first correction to $U_k$ taking into account the momentum dependence of the correlation functions $\Gamma^{(n)}_k(q_1,\dots,q_n)$.  $Z_k(\rho)$ 
enables to compute the anomalous dimension of the field $M$. Of course, the ansatz (\ref{ansatz}) cannot be an exact solution of
$\Gamma_k[M]$ and thus $U_k(\rho)$ and $Z_k(\rho)$ must be defined in such a way that their flows do not get any contribution from  the $O(\nabla^{2n})$ 
terms with $n>1$ that are supposed to be neglected in the ansatz. It is convenient to define them as:
\begin{Eqnarray}
U_k(\rho)&=& \frac{\Gamma(\rho)}{V}\\
Z_k(\rho)&=&\frac{1}{V} \frac{d}{dq^2}\Gamma^{(2)}_k(q_,-q)(\rho)_{\vert_{q^2=0}}
\end{Eqnarray}
where $V$ is the volume of the system and $\rho$ is a uniform, that is $x$-independent, field configuration. Inserting these definitions in
the RG flow equation (\ref{rgexact}) leads to the RG equation for $U_k(\rho)$ and $Z_k(\rho)$. These equations are partial 
differential equations.
It is interesting to notice that it is possible to further truncate $\Gamma_k[M]$ while preserving  qualitatively and semi-quantitatively 
the properties of the method. This additional truncation consists in  field expanding  $Z_k(\rho)$  and/or  the potential $U_k(\rho)$, 
that is\cite{tetradis94,berges02}:

1)  $U_k(\rho)$ is truncated by keeping only a polynomial in $\rho$ of low degree. It is convenient (and more accurate) in this case to expand
$U_k(\rho)$ around its running minimum $\kappa_k$. The simplest non trivial field truncation is: $U_k(\rho)=\lambda_k/2(\rho- \kappa_k)^2 + O((\rho-\kappa_k)^3)$.

2)  $Z_k(\rho)$ is truncated by keeping only a field-independent coefficient: $Z_k(\rho)\to Z_k \equiv Z_k(\rho=\kappa_k)$. 
This is the simplest truncation that allows to compute a non trivial anomalous dimension. Let us notice that it has for long been 
believed that the anomalous dimension was essentially arbitrary within the derivative expansion and that it was therefore useless
to include a  $Z_k(\rho)\neq 1$. Taking  $Z_k(\rho)= 1$ --- and thus $\eta=0$ --- is called the Local Potential Approximation (LPA) and was 
first used (in a slightly different way) by Wegner and Houghton back in the seventies\cite{wegner73}.

If we combine  truncations 1 and 2,
the RG flow becomes a set of three simple coupled differential equations for $\kappa_k, \lambda_k$ and $Z_k$\cite{tetradis94}. A very important point 
with this truncation when it
is implemented on the $O(N)$ models is that it allows to recover i) the one-loop result obtained perturbatively in $d=4-\epsilon$, ii)
 the $N\to\infty$ result in any dimension and, for $N\geq 3$, iii) the one-loop result obtained in $d=2+\epsilon$ from the non linear 
sigma model\cite{berges02,delamotte04}. 
This is very different from the perturbative results that are very difficult to extrapolate beyond their domain of applicability.
Let us also notice that for $N=2$ and $d=2$, this truncation enables to retrieve most of the qualitative as well as 
some quantitative features of the Kosterlitz-Thouless transition\cite{grater95}.

There are two origins 
for the errors made with an ansatz such as Eq.(\ref{ansatz}) supplemented or not by a field expansion of $U_k(\rho)$ and/or $Z_k(\rho)$. 
First, the exact RG equations for  $U_k$ and $Z_k$ involve the neglected  terms of the derivative expansion of order $\nabla^4,\nabla^6, \dots$. 
Moreover, if a field expansion of $U_k(\rho)$ is performed as in truncation 1) above, the RG equations of  $\kappa_k$ and $\lambda_k$ involve the 
coupling constants of the next terms of the expansion, that is the coefficients of $(\rho-\kappa_k)^n$ with $n>2$. Second, the choice 
of a cut-off function $R_k$ that, in principle, has no influence since $R_k(q^2)$ vanishes identically in the limit $k\to 0$, 
does matter once truncations are performed. Many studies have been devoted to finding an optimal 
choice\cite{liao00,litim02,canet03a,canet03b,canet05}. None of them 
gives a complete solution to this problem.

Let us now work out the RG equations  obtained with a standard truncation as (\ref{ansatz})  for the effective 
potential and anomalous dimension.
 Let us first recall that since at a second order phase transition the correlation length is infinite, 
the system becomes scale invariant at the transition point (at sufficiently large
distance compared with the lattice spacing). Thus, in units of the (inverse) running block spin size $k$, 
it becomes invariant under the RG transformations (for $k\ll\Lambda$). It is therefore convenient to work with 
dimensionless quantities: in terms of them the transition corresponds to a fixed point of the RG flow. Actually, 
to find out a fixed point, it is  also necessary to work with a renormalized magnetization. Indeed, 
at the transition point,
the magnitude of the running (spontaneous) magnetization $M$ never stops flowing towards 0 as $k\to 0$, which
 a priori  prevents from finding a fixed point.
 It is thus necessary to work with a (dimensionless) renormalized magnetization whose 
magnitude tends to a finite value as $k\to 0$. We thus define
\begin{Eqnarray}
\tilde{\rho}&=& Z_k k^{2-d} \rho\\
 u_k(\tilde{\rho}) &=& k^{-d} U_k(\rho).
\end{Eqnarray}
With truncation 2) above and with the regulator of Eq.(\ref{regulateur}), the RG equations 
for the potential $u_k(\tilde{\rho})$ and for $Z_k$ become very simple:
\begin{Eqnarray}
\p_s u_k(\tilde{\rho})&&= -d u_k(\tilde{\rho}) +(d-2+\eta_k ) \tilde{\rho} u_k'(\tilde{\rho})\nonumber \\
&&+ \frac{4 v_d}{d}\left(1-\frac{\eta_k}{d+2}\right)\frac{1}{1+u_k'(\tilde{\rho}) 
+2\tilde{\rho} u_k''(\tilde{\rho}) }\ \ \ \ \ \ \ \ \\
\eta_k&=&-\p_s\log Z_k= \frac{16 v_d}{d} \frac{\lambda_k \kappa_k}{1+2\lambda_k \kappa_k },
\end{Eqnarray}
where $v_d=1/(2^{d+1}\pi^{d/2} \Gamma(d/2))$,  primes denote  derivatives  with respect to $\tilde{\rho}$, $\kappa_k$ is
the minimum  of $u_k(\tilde{\rho})$ and $\lambda_k=u_k''(\kappa_k)$. $\eta_k$ is the running anomalous dimension which 
goes to a fixed point value as $k\to 0$ --- the physical  anomalous dimension --- if the potential itself reaches a fixed point solution.

This theoretical framework --- or those deriving from more sophisticated truncations --- have turned out to be very
powerful. We show in the following some results that have been obtained for the Ising and $O(N)$ models by ourselves 
and by other groups. However, before this, let us show how to adapt this formalism for out of equilibrium statistical systems.

\subsection{The out of equilibrium case}

For out of equilibrium systems the very notion of free energy does no longer exist in general. However, it is still possible
to define  generating functionals of  correlation functions, analogous to the partition function ${\cal Z}$. 
Moreover, even if it is no longer possible to speak of Boltzmann 
weights and hamiltonians, the generating functionals ${\cal Z}$ are still often given by  functional integrals of 
weights that, formally, enable to define ``hamiltonians'',  called actions in this context. The Legendre transform 
$\Gamma$ of $\log {\cal Z}$
is then analogous to the Gibbs free energy and is the generator of the 1PI correlation functions. There are two main means to 
build field theories in out of equilibrium statistical mechanics: either from a Langevin formulation of the problem or
from a more microscopic approach --- if it exists --- {\it \`a la} Doi-Peliti\cite{doi76,peliti84}, that is from a master equation.

 Even if 
approaches at and out of equilibrium are close together, some physical differences are crucial, such as the irreversibility of the dynamics, 
the violation of the 
fluctuation-dissipation theorem, etc. Technically, these differences have a translation in the formalism: i) the field 
theories are written not only in terms of the ``physical'' field(s) but
also in terms of the response field(s), ii) temporal fluctuations  are taken into account 
by a single time derivative (contrary to the Laplacian for spatial fluctuations), iii) the actions are not ``hermitic'', that
is, the physical and response fields do not necessarily play symmetric roles\cite{cardy96c}. However, all these differences do not prevent
from constructing an analog of the effective average action method for out of equilibrium systems\cite{canet04a,canet04b}. The non trivial point is to
choose a $R_k$ function. Should we take it space and time dependent? acting only on the physical field or on both the physical and
response fields? Since the free field theory corresponds to the following theory:
\begin{Eqnarray}
{\cal Z}&=& \int D\phi D\bphi e^{-{\cal S}_0}\\
\hbox{with}&&\\
{\cal S}_0&=&\int d^dx dt\; \bphi(x,t) \left(\p_t - D \nabla^2\right)\phi(x,t)
\end{Eqnarray}
it is natural to define the $R_k$ term by:
\be
\Delta {\cal S}_k = \int d^dx d^dy dt dt'\; \bphi(x,t)R_k(x-y,t-t')\phi(y,t').
\ee
We have also chosen to take it time-independent so that the time integral  in the RG equation can
be trivially performed\cite{canet04a}. It is not known whether this corresponds to a ``good'' choice from the accuracy point of
view. Once these choices have been made, it is possible for a non trivial model with action ${\cal S}$ to define:
\be
{\cal Z}_k[J,\bar{J}] = \int \: {\cal D}\phi {\cal D}\bphi\;  e^{
-{\cal S}[\phi,\bphi]- \Delta {\cal S}_k[\phi,\bphi] +\int J.\phi +\int \bar{J}.\bphi } .
\label{zk_hors_eq}
\ee
The effective average action is defined as in Eq.(\ref{legendre}) by:
\begin{Eqnarray}
\Gamma_k[\psi,\bpsi] + {\cal W}_k[J,\bar{J}]&=& \int J \psi + \int \bar{J}\bpsi \\
                                   &&\ \    - \int_q R_k(q^2) \psi_{q,\omega} \bpsi_{-q,-\omega}
\label{legendre_hors_eq}
\end{Eqnarray}
with
\be
\psi (x,t)= \frac{\delta {\cal W}_k}{\delta J(x,t)}\ \ \ \hbox{and}\ \ \ \bpsi(x,t)= \frac{\delta {\cal W}_k}{\delta \bar{J}(x,t)}.
\label{psi_bpsi}
\ee
It is convenient to define both the 2$\times$2 matrix $\hat{\Gamma}_k^{(2)}[\psi,\bpsi]$ of second functional derivatives of $\Gamma_k[\psi,\bpsi]$ 
with respect  to $\psi$ and $\bpsi$ and the 2$\times$2 matrix $\hat{R}_k$, the off-diagonal elements of which are $R_k(q^2)$.
Then, the out of equilibrium NPRG equation writes\cite{canet04a}:
\be
\p_s\Gamma_k = \half \Tr \int_{q,\omega} \p_s\hat{R}_k \left(\hat{\Gamma}^{(2)}_k(q,\omega;-q,-\omega)+\hat{R}_k(q^2)\right)^{-1}.
\label{rgexact_hors_eq}
\ee

A typical ansatz for a reaction-diffusion problem is:
\be
\Gamma_k[\psi,\bpsi]= \int d^dxdt\; \left\{U_k(\psi,\bpsi)+  Z_k\bpsi\left(\p_t-D_k\nabla^2\right) \psi  \right\}.
\label{ansatz_hors_eq}
\ee
Within this formalism some problems of branching and annihilating random walks (BARW) have been addressed\cite{canet04b}
that we briefly sketch further on.

\section{Some results obtained with the NPRG method}

Among others, a crucial problem as for the NPRG method combined with the derivative expansion, is to estimate
the quantitative accuracy of the results obtained. More precisely, it
would be highly desirable to be able to estimate  error bars. As usual, this is a very difficult
task when no analytic result is known on the behavior of the series encountered in the method. 
As already mentioned, this question is closely related to the choice of the cut-off function $R_k$. 

\begin{table}[htbp]
\begin{center}
\begin{tabular}{|c |  c c  |}
\hline 
  \hspace{0.3cm} order  \hspace{0.3cm}    & $\nu$ & $\eta$   \\
\hline 
   $\p^0$   &   0.6506 &  0 \\ 
   $\p^2$   & 0.6281   &  0.044\\
   $\p^4$   & 0.632    &  0.033 \\
\hline 
7-loops  & 0.6304(13) & 0.0335(25)\\
\hline
\end{tabular}
\caption{Critical exponents of the three dimensional Ising model.  $\p^0$, $\p^2$ and $\p^4$  correspond to the order of the truncation of the 
derivative expansion (NPRG method)\cite{canet03b}. For completeness, we have recalled in the last line, the results obtained 
perturbatively\cite{zinnjustin89}.}
\label{tableI}
\end{center}
\end{table}

We have studied the critical exponents of the Ising model in three dimensions as a testing ground of the convergence of
the derivative expansion at order $\nabla^0$ (LPA), $\nabla^2$ and $\nabla^4$\cite{canet03b}.
The rule of thumb adopted consists in evaluating the error bar through the evolution of the values of the exponents with the order
of the truncation. In the different  cases studied, we have found that the distance between the best known results 
and ours decreases when increasing the order of the truncation of the derivative expansion. Thus, it seems 
reasonable to believe that the error made at a given order of the derivative expansion is given by the distance
between the results of the last two orders. More precisely, at each order of the truncation, we have studied 
the dependence of the critical exponents on the choice of function $R_k$. Each time, we have tried to find the 
best $R_k$ by requiring  
the exponents to depend as weakly as possible on the choice of $R_k$ (since they verify this property in the exact theory). We have 
therefore considered a one-parameter family of functions $R_k$
and have computed the critical exponents for all elements of this family. The function $R_k$ that is selected as  optimal
is that for which the exponents are stationary with respect to a change of $R_k$ (this is an 
implementation of the Principle of Minimal Sensitivity)\cite{canet03a}. At each order of the derivative expansion, this leads to a
set of optimal exponents. At order $\nabla^4$, these optimal exponents are quite close to the best known values, see Table I. 
Thus, although only three orders of the derivative expansion are known, it seems that, at least in the three dimensional Ising case,
the derivative expansion converges rapidly. It is important to notice that no resummation  of any kind has been necessary
to obtain these results.

As a conclusion, one can see on Table I that the results obtained from the NPRG method are in a fairly 
good quantitative agreement with the best known results
for the three dimensional Ising model. This is, of course, very encouraging as for the reliability of the method and the convergence
of the derivative expansion. Let us moreover emphasize that the Ising model belongs to the very small class of models, the perturbation
expansion of which involves only one coupling constant and is known to be Borel summable\cite{zinnjustin89}. In many cases, things 
are much more complicated.
A famous example is given by the three dimensional frustrated spin systems, either the triangular antiferromagnets or the helimagnets. 
Here, the results obtained by means of the resummed perturbation series and those obtained from the NPRG are qualitatively 
different\cite{tissier00,pelissetto01a,pelissetto02,tissier03,delamotte04,calabrese04}.
The former predicts a second order phase transition whereas the latter predicts  very weakly first order transitions. The debate is still open
and it will be interesting to see which prediction is correct.

Let us now turn to out of equilibrium systems and more precisely to branching and annihilating random walks. The problem is the following.
On a $d$-dimensional lattice, particles are diffusing (with rate $D$) and can create offsprings through the reaction $A\to A+A$,
 with rate $\sigma$. When they meet on the same site, 
they can annihilate through  $A+A\to \emptyset $  with rate $\lambda$. The problem is to determine whether the system undergoes a continuous phase transition between
an active and an inactive (absorbing) phase and to which universality class it belongs. The field theory is obtained  from the master
equation and the action reads\cite{cardy98}:
\begin{Eqnarray}
{\cal S}= \int d^dx\; dt \left\{ \bphi \left(\p_t -D \nabla^2 \right)\phi\right. &+&\sqrt{2 \sigma \lambda } (\bphi \phi^2 -\bphi^2 \phi)\nonumber \\
&+& \left.\lambda \bphi^2\phi^2  \right\}.
\label{actionbarw}
\end{Eqnarray}

 First of all, a mean field study predicts that
the system is always in the active phase for $\sigma\neq 0$. From a perturbative analysis performed at and  below $d=2$, Cardy and T\"auber 
have obtained the phase diagram for small $\lambda/D$ and $\sigma/D$, showing unambigously that fluctuations are able to destabilize the active
phase so that an absorbing phase exists
at finite $\sigma/D$ for $d\leq2$\cite{cardy98}. As for the universality class, this phase transition belongs to the Directed Percolation one. However,
their calculations also suggest that for all  dimensions larger than two, the fluctuations are not strong enough to 
destabilize the mean field
result and that, therefore, the system is always in the active phase. Using the NPRG method, we have re-examined the physics of this system.
We have been able to compute the critical exponents in all dimensions and have shown that the phase diagram obtained perturbatively is 
wrong\cite{canet04b}. More precisely, 
above dimension two, the system is indeed always in the active phase at small $\sigma/D$ and  $\lambda/D$ as predicted by the perturbative analysis. But there
 is a threshold value of $\lambda/D$ above which an absorbing phase can exist. By numerically integrating the RG flow  from dimensions 3 to 6
we have determined these threshold values as well as the complete phase diagram\cite{canet04b}. 
\begin{figure}[ht]
\includegraphics[height=86mm,angle=-90]{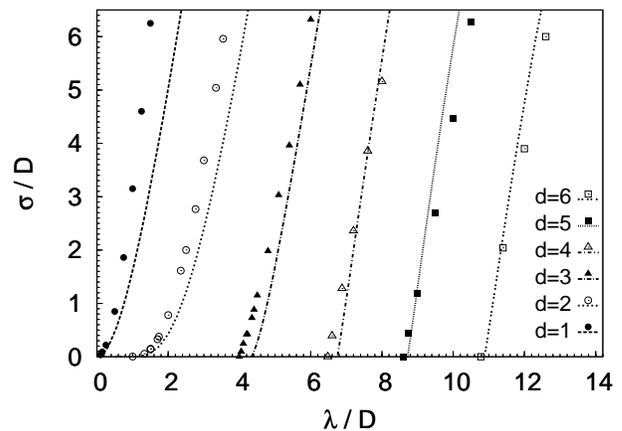}
\caption{Phase diagrams of BARW 
$A \xrightarrow{\sigma} 2A$, $2A\xrightarrow{\lambda}\varnothing$
in dimensions 1 to 6. Lines present NPRG results, rescaled as explained 
in the text. Symbols follow from numerical simulations. 
For each dimension, the active phase lies on the left of the transition 
line, the absorbing phase on the right.}
\label{figure1}
\end{figure}
In  Figure \ref{figure1} is displayed both the
NPRG results and those obtained from  numerical simulations. Only one free parameter has been used to match the numerical and analytical results.
This parameter takes into account the fact that both approaches differ at the microscopic scale $\Lambda$  since one is defined on the lattice 
and the other in the continuum. However, it is  known that both models should match after some RG iterations since, at sufficiently large scale, 
the lattice model becomes homogeneous and isotropic and  therefore can be described by a continuum theory. Thus, after a proper matching of the
ultraviolet scales $\Lambda$ of both models ---  requiring one rescaling parameter --- the 
results can be compared quantitatively. The agreement is excellent and proves that, also in this example, the NPRG method is
reliable. Let us emphasize that  the phase diagram is a non universal feature of the model. Computing it requires
to keep track all along the RG flow of both  relevant and  irrelevant couplings since they all contribute to the non universal
behavior of the model. While this is in general a formidable task in perturbation theory, that anyway is useless if these
couplings are large, it just consists in the NPRG approach in integrating the flow equations for the potential with the bare potential 
of Eq.(\ref{actionbarw}) as initial condition at scale $\Lambda$\cite{parola95,seide99,canet04b}.

\section{Conclusion and some open problems}
Let us now come to a conclusion. The NPRG method seems to be a promising way to tackle  the strong coupling regime of (at least some) field theories while
avoiding problematic resummation methods used in perturbative renormalization. It also enables to compute non universal quantities
such as phase diagrams because it can handle the functional nature of the effective potential with 
both its relevant and  irrelevant couplings, that are most of the time infinitely many\cite{seide99,canet04b}. For the reaction-diffusion problem  
we have discussed in this article, the NPRG has allowed us 
to find that a phase transition exists in all dimensions while perturbative renormalization failed, probably even in  a all-order analysis. 
This example (and it is not unique, see for instance\cite{morris95a,berges96,gersdorff01,rosa01,delamotte04,canet04b}) 
already shows that this method leads to highly non trivial results. 
Now, an interesting question is: can we classify
problems for which perturbative theory could fail whereas this method could still be relevant and, reciprocally, can we list its weak points? 
Let us try.

We have already mentioned the problems encountered in the three dimensional frustrated systems for which the NPRG and the six-loop perturbative
computations lead to different results\cite{delamotte04}. It is interesting to understand why, since this problem probably belongs to a whole class of problems 
where perturbative theory is problematic and where the NPRG method could be fruitful (see \cite{delamotte04}). The problem is the following. The Ginzburg - 
Landau - Wilson field theory ($\phi^4$-like) relevant for the description of the critical physics of a system is determined by three properties of
the system: i) the symmetry breaking scheme, $G \to H$, between the high ($G$) and the low ($H$) temperature phases, ii) the number $N$ 
of components of the order parameter and iii) the space dimension $d$. The number of coupling constants in the hamiltonian depends on the 
symmetry group $G$ and the
lower this symmetry, the larger the number of  invariant couplings in the hamiltonian. For all systems with a symmetry
lower than $O(N)$, at least two coupling constants are present in the hamiltonian. In these cases, the number of fixed points is larger than one,
contrary to the $O(N)$ case where there is only one.  When $N$ and/or $d$ are varied, the fixed points move in the coupling constant
space and it is generically observed that, at fixed $d$, there exists a critical value $N_c$ of $N$ where two of them collapse. Two possible 
situations can then occur: either both  fixed points disappear when $N$ is further varied (they become complex) or they exchange their stability.
 These two situations are, for instance, encountered  in  frustrated 
systems and in  ferromagnetic systems with cubic anisotropy. The difficulty is that as $d$ is varied, $N_c$ changes and, of course, the
problem is to compute  $N_c(d)$ since it is crucial to determine if the fixed points exist in $d=3$ for  $N=1,2,3$ and, if they do, 
which one is stable. This is a highly non trivial task in perturbation theory that requires, in the known cases, resummations of the series
 obtained at least at  four loops. Clearly, if the NPRG method leads to quantitatively reliable results in these cases, this would avoid
 tedious perturbative calculations, that are anyway useless if more than six loop results are necessary to obtain a converged
value of $N_c(d=3)$. These problems belong to the whole class of ``strong coupling problems'' of field theory.

We have already mentioned another class of problems where NPRG could be useful: those where perturbation theory fails at all orders to describe
the critical physics. An important example is given by the RFO(N)M where dimensional reduction has been proven to be wrong\cite{imbrie84,bricmont87}. 
Precisely on this example,
the NPRG approach has been used very recently\cite{tarjus05}. The origin of the problem is that the effective potential $u_k$ acquires non analytic 
contributions at finite $k$ that are missed by perturbation theory. These non analyticities have been successfully computed from 
the NPRG method since the functional nature of the effective potential can be dealt with. It will be very interesting to see if this method
is able to obtain qualitatively and quantitatively correct results for the Ising case in three dimensions. Anyhow, the present results in this model
already open the way to a  renormalization group which is both functional and non perturbative.

A third instance where NPRG could turn out to be useful is the study of phase transitions where topological defects play a role 
(especially in low dimensions).
Let us recall that  almost nothing is known in the physics of phase transitions induced by vortices in two dimensions apart from
the Kosterlitz-Thouless transition for which the decoupling between  spin waves and vortices allows a perturbative study of the transition.
It could seem strange at first sight that, in situations where non trivial spatial configurations of the field are crucial, anything relevant
 can be learnt from a derivative expansion that keeps only the lowest order(s) in the momentum  dependence of 
correlation functions. Let us
however make two general remarks. First, the Kosterlitz-Thouless transition has been studied by the NPRG method {\it without} introducing 
the vortices by hand {\it \`a la } Villain. Almost all the physics is quantitatively recovered apart from the fact that the line of fixed points is 
replaced by a line of quasi-fixed points where the $\beta$-function of the temperature is very small although not strictly vanishing\cite{gersdorff01}. Second,
spatially non trivial field configurations can play a great role in the summation over fluctuations, that is  for the computation of the 
functional integral, while correlation functions, that are the results of this summation, can be very smooth functions of the momenta. Thus, it
is possible that a derivative expansion succeeds in capturing the physics of topological defects. This would be extremely interesting
since in many cases other than the $O(2)$ case, no Villain trick is available to write down a field theory that can
be studied perturbatively.

Let us now review some of the drawbacks of the NPRG method that would clearly require much effort to be overcome. First, the derivative expansion
forbids to compute the momentum dependence of correlation functions. It just focuses on the flow of coupling constants with emphasis on those
of the effective potential. It is therefore insufficient to compute cross sections in particle physics or structure factors. Second, it is not
easy to point out the ``small parameter'' that underlies the validity of the derivative expansion. We can postulate that the larger
the anomalous dimension, the worse its convergence. 
However,  this important assumption has yet not been supported by a systematic theoretical analysis.
Third, the very way the cut-off function $R_k(q^2)$ is introduced can be problematic. For instance, it breaks gauge 
symmetry and it is clearly one of the big challenge of the NPRG to allow gauge invariant computation\cite{arnone03}. Fourth, the choice of an ansatz
for $\Gamma_k$ relies on the choice of its field content. In statistical mechanics, contrary to particle physics,
the microscopic degrees of freedom, at scale $\Lambda$, are known in general and the difficulty is to propose an ansatz in terms of 
the low energy degrees of freedom. As already emphasized, they can qualitatively differ from those at high energy
 --- e.g. they are bound states --- and it is non trivial to know whether the RG flow of, say, a potential will signal this change. The problem
is that, although the physics can still be described in terms of the microscopic degrees of freedom, 
it can then turn out to be very complicated.
 For instance, while the physics is ``local'' in terms of the bound state field, it can become non local in terms of the microscopic
degrees of freedom. Thus, a derivative expansion formulated in terms of the bound state field can be accurate whereas it is not in terms
of the original fields\cite{gies02}. The choice of low energy degrees of freedom is therefore crucial for devicing 
an ansatz for $\Gamma_k$ and it is not clear
up to now if there exists a systematic way for, at least, detecting a bad choice. 

NPRG is still in its infancy. For sure, some of the above mentioned problems will be solved in the future. Some others will remain out of reach.
As usual, numerous physical problems will challenge us and will force us to better understand the NPRG method  as well as its intrinsic limits.

\section{Acknowledgements}
We thank Y. Holovatch who gave us the opportunity to contribute to the Feschrift dedicated to the 60th birthday of R. Folk.
We are grateful to D. Mouhanna, M. Tissier, J. Vidal and N. Wschebor with whom parts of the results that have been shown in this 
article have been obtained. The numerous discussions we have had with them have been crucial for our understanding of the NPRG. The LPTL is unit\'e 
mixte du CNRS UMR 7600  and  L.C. acknowledges financial support by the European Community's Human Potential Programme under contract 
HPRN-CT-2002-00307, DYGLAGEMEM.


\begin{thebibliography}{10}

\bibitem{lebellac92}
M. {Le Bellac}, {\em Quantum and statistical field theory} (Oxford University
  Press, Oxford, 1992).

\bibitem{goldenfeld92}
N. Goldenfeld, {\em Lectures on Phase Transitions and the Renormalization
  Group} (Addison-Wesley, Reading, MA, 1992).

\bibitem{itzykson89}
C. Itzykson and J. Drouffe, {\em Statistical Field Theory} (Cambridge
  University Press, Cambridge, England, 1989).

\bibitem{zinnjustin89}
J. Zinn-Justin, {\em Quantum Field Theory and Critical Phenomena}, 3 ed.
  (Oxford University Press, New York, 1989).

\bibitem{pelissetto00a}
A. Pelissetto and E. Vicari, Phys. Rev. B {\bf 62},  6393  (2000).

\bibitem{holovatch02}
{{Yu.} Holovatch}, V. Blavats'ka, M. Dudka, C.~V. Ferber, R. Folk, and T.
  Yavors'kii, Int. J. Mod. Phys. B {\bf 16},  4027  (2002).

\bibitem{folk03}
R. Folk, {{Yu.} Holovatch}, and T. Yavors'kii, Phys.-Usp. {\bf 173},  175
  (2003).

\bibitem{tissier00}
M. Tissier, B. Delamotte, and D. Mouhanna, Phys. Rev. Lett. {\bf 84},  5208
  (2000).

\bibitem{pelissetto01a}
A. Pelissetto, P. Rossi, and E. Vicari, Phys. Rev. B {\bf 63},  140414  (2001).

\bibitem{pelissetto02}
A. Pelissetto, P. Rossi, and E. Vicari, Phys.Rev. B {\bf 65},  020403  (2002).

\bibitem{tissier03}
M. Tissier, B. Delamotte, and D. Mouhanna, Phys. Rev .B {\bf 67},  134422
  (2003).

\bibitem{delamotte04}
B. Delamotte, D. Mouhanna, and M. Tissier, Phys. Rev. B {\bf 69},  134413
  (2004).

\bibitem{calabrese04}
P. Calabrese, P. Parruccini, A. Pelissetto, and E. Vicari, cond-mat/0405667
  (2004).

\bibitem{frey94}
E. Frey and U. Ta{\"{u}}ber, Phys. Rev. E {\bf 50},  1024  (1994).

\bibitem{wiese98}
K.~J. Wiese, J. Stat. Phys. {\bf 93},  143  (1998).

\bibitem{parisi79}
G. Parisi and N. Sourlas, Phys. Rev. Lett. {\bf 43},  744  (1979).

\bibitem{imbrie84}
J.~Z. Imbrie, Phys. Rev. Lett. {\bf 53},  1747  (1984).

\bibitem{bricmont87}
J. Bricmont and A. Kupianen, Phys. Rev. Lett. {\bf 59},  1829  (1987).

\bibitem{tarjus05}
G. Tarjus and M. Tissier, cond-mat/0410118  (2005).

\bibitem{brezin01}
E. Br\'ezin and {C. {De} Dominicis}, Eur. Phys. J. B {\bf 19},  467  (2001).

\bibitem{parisi02}
G. Parisi and N. Sourlas, Phys. Rev. Lett. {\bf 89},  257204  (2002).

\bibitem{wilson74}
K.~G. Wilson and J. Kogut, Phys. Rep. C {\bf 12},  75  (1974).

\bibitem{polchinski84}
J. Polchinski, Nucl. Phys. B {\bf 231},  269  (1984).

\bibitem{wetterich93c}
C. Wetterich, Phys. Lett. B {\bf 301},  90  (1993).

\bibitem{tetradis94}
N. Tetradis and C. Wetterich, Nucl. Phys. B [FS] {\bf 422},  541  (1994).

\bibitem{ellwanger94c}
U. Ellwanger, Z. Phys. C {\bf 62},  503  (1994).

\bibitem{morris94a}
T.~R. Morris, Int. J. Mod. Phys. A {\bf 9},  2411  (1994).

\bibitem{morris94b}
T.~R. Morris, Phys. Lett. B {\bf 329},  241  (1994).

\bibitem{berges02}
J. Berges, N. Tetradis, and C. Wetterich, Phys. Rep. {\bf 363},  223  (2002).

\bibitem{litim01}
D. Litim, Phys. Rev. D {\bf 64},  105007  (2001).

\bibitem{wegner73}
F.~J. Wegner and A. Houghton, Phys. Rev. A {\bf 8},  401  (1973).

\bibitem{grater95}
M. Gr\"ater and C. Wetterich, Phys. Rev. Lett. {\bf 75},  378  (1995).

\bibitem{liao00}
S. Liao, J. Polonyi, and M. Strickland, Nucl. Phys. B {\bf 567},  493  (2000).

\bibitem{litim02}
D.~F. Litim, Nucl. Phys. B {\bf 631},  128  (2002).

\bibitem{canet03a}
L. Canet, B. Delamotte, D. Mouhanna, and J. Vidal, Phys. Rev. D {\bf 67},
  065004  (2003).

\bibitem{canet03b}
L. Canet, B. Delamotte, D. Mouhanna, and J. Vidal, Phys. Rev. B {\bf 68},
  064421  (2003).

\bibitem{canet05}
L. Canet, hep-th/0409300  .

\bibitem{doi76}
M. Doi, J. Phys. A {\bf 9},  1479  (1976).

\bibitem{peliti84}
L. Peliti, J. Phys. (Paris) {\bf 46},  1469  (1984).

\bibitem{cardy96c}
J.~L. Cardy,  in {\em Proceedings of Mathematical Beauty of Physics}, edited by
  J.-B. Zuber (Adv. Ser. in Math. Phys., vol. 24 p. 4780, 1996).

\bibitem{canet04a}
L. Canet, B. Delamotte, O. Deloubri\`ere, and N. Wschebor, Phys. Rev. Lett.
  {\bf 92},  195703  (2004).

\bibitem{canet04b}
L. Canet, H. Chat\'e, and B. Delamotte, Phys. Rev. Lett. {\bf 92},  255703
  (2004).

\bibitem{cardy98}
J.~L. Cardy and U.~C. Ta{\"{u}}ber, J. Stat. Phys. {\bf 90},  1  (1998).

\bibitem{parola95}
A. Parola and L. Reatto, Adv. Phys. {\bf 44},  211  (1995).

\bibitem{seide99}
S. Seide and C. Wetterich, Nucl. Phys. B {\bf 562},  524  (1999).

\bibitem{morris95a}
T.~R. Morris, Nucl. Phys. Proc. Suppl. {\bf 42},  811  (1995).

\bibitem{berges96}
J. Berges, N. Tetradis, and C. Wetterich, Phys. Rev. Lett. {\bf 77},  873
  (1996).

\bibitem{gersdorff01}
G. v.~Gersdorff and C. Wetterich, Phys. Rev. B {\bf 64},  054513  (2001).

\bibitem{rosa01}
L. Rosa, P. Vitale, and C. Wetterich, Phys. Rev. Lett. {\bf 86},  958  (2001).

\bibitem{arnone03}
S. Arnone, A. Gatti, and T. Morris, Phys. Rev. D {\bf 67},  085003  (2003).

\bibitem{gies02}
H. Gies and C. Wetterich, Phys. Rev. D {\bf 65},  065001  (2002).

\end{thebibliography}

\end{document}